\documentclass[prl,twocolumn,superscriptaddress]{revtex4} 
\usepackage{epsfig}

\usepackage{graphicx}

\usepackage{color,colordvi,graphicx,pstcol}
\usepackage{amsmath}
\usepackage{graphicx}
\usepackage{dcolumn}
\usepackage{bm}

\newcommand{\hide}[1]{}

\newcommand{\eq}[1]{Eq.\,(\ref{#1})}
\newcommand{\eqs}[1]{Eqs.\,(\ref{#1})}

\newcommand{\fig}[1]{Fig.\,\ref{#1}}

\newcommand{\nofig}[1]{\ref{#1}}

\newcommand{\Cc}{{\cal C}}

\newcommand{\Nc}{{\cal N}}
\newcommand{\e}{{\rm e}}

\newcommand{\s}{{\sigma}}
\newcommand{\Gm}{\Gamma}

\newcommand{\gm}{\gamma}

\newcommand{\cum}[1]{\left\langle\!\left\langle #1 \right\rangle\!\right\rangle}
\newcommand{\comm}[2]{\left[ #1 , #2 \right]}
\newcommand{\erw}[1]{\left\langle #1 \right\rangle}

\newcommand{\ket}[1]{\ensuremath{\left| #1 \right\rangle}}
\newcommand{\bra}[1]{\ensuremath{\left\langle #1 \right|}}

\newcommand{\To}{\ensuremath{\rightarrow}}

\newcommand{\beq}{\begin{equation}}
\newcommand{\eeq}{\end{equation}}

\begin{document}
\title{Superradiance in ultracold Rydberg gases}
\author{T. Wang}
\affiliation{Department Of Physics, University of Connecticut,
Storrs, CT 06269}

\author{S. F. Yelin}
\affiliation{Department Of Physics, University of Connecticut,
Storrs, CT 06269} \affiliation{ITAMP, Harvard-Smithsonian Center
for Astrophysics, Cambridge, MA 02138}

\author{R. C\^{o}t\'{e}}
\affiliation{Department Of Physics, University of Connecticut,
Storrs, CT 06269}

\author{E. E. Eyler}
\affiliation{Department Of Physics, University of Connecticut,
Storrs, CT 06269}

\author{S. M. Farooqi}
\affiliation{Department Of Physics, University of Connecticut,
Storrs, CT 06269}

\author{P. L. Gould}
\affiliation{Department Of Physics, University of Connecticut,
Storrs, CT 06269}

\author{M. Ko\v{s}trun}
\affiliation{Department Of Physics, University of Connecticut,
Storrs, CT 06269} \affiliation{ITAMP, Harvard-Smithsonian Center
for Astrophysics, Cambridge, MA 02138}

\author{D. Tong}
\affiliation{Department Of Physics, University of Connecticut,
Storrs, CT 06269}

\author{D. Vrinceanu }
\affiliation{Theoretical Division, Los Alamos National Laboratory, NM 87545}

\date{\today}
\begin{abstract}

Experiments in dense, ultracold gases of rubidium Rydberg
atoms show a considerable decrease of the radiative excited state
lifetimes compared to dilute gases. This accelerated decay
is explained by collective and cooperative effects, leading
to superradiance. A novel formalism to calculate effective
decay times in a dense Rydberg gas shows that for these atoms
the decay into nearby levels increases by up to three
orders of magnitude. Excellent agreement between theory and
experiment follows from this treatment of Rydberg decay behavior.

\end{abstract}

\maketitle

In recent years, ultracold atomic gases have been used to probe a
variety of many-body phenomena such as Bose-Einstein condensation
~\cite{bec1,bec2} and degenerate Fermi gases ~\cite{FermiGas}. In
addition to collective effects due to particle statistics,
other manifestations of many-body physics have been explored, such
as in slow-light experiments ~\cite{ObservationStorage} and in
ultracold Rydberg gases (e.g. the diffusion of excitations through
resonant collisions ~\cite{pillet98} and the blockade mechanism
~\cite{BlockadeExp}). Another important fundamental collective
effect is superradiance, in which photon exchange between atoms
modifies the behavior of the sample. In particular, {\em
cooperative} effects due to virtual photon exchange can lead to
the formation of so called Dicke states \cite{Dicke}. These states
are the symmetric superposition of all states with the same total
excitation level for constant atom number $N$. Interest in Dicke
states has grown recently because of their potential advantages in
quantum information processing~\cite{MesoSpin} and their
importance in the behavior of Bose-Einstein
condensates~\cite{SuperBEC}.

In this Letter, we are interested in many-body physics involving
photon exchange in an ultracold gas of Rydberg atoms. Because
superradiance depends on the atomic density per cubic wavelength,
and because radiative decay of Rydberg atoms takes place
predominantly between the closely spaced upper levels,
ultracold Rydberg gases are ideal systems to study
superradiance. In fact, Rydberg atoms have many interesting
properties: their size can become comparable to the atomic
separation, and they have huge dipole moments $\wp\sim n^2$,
where $n$ is the principal quantum number of the Rydberg state.
In addition, for long-wavelength transitions between neighboring
states of high $n$ the ``cooperative parameter''
$\Cc=\Nc \lambda^3/4\pi^2$ (where $\Nc$ is the density of
atoms, $\lambda$ is the transition wavelength), is large for Rydberg
atoms, which means collective effects are much easier to obtain
than for ground-state atoms~\cite{DelayCoLength}. This was
confirmed in earlier experiments for Rydberg atoms at high \cite{RydSuper2,RydSuper3} and low temperatures \cite{raithel}. Note that these many-body
effects may pose a limit on the measurement of lifetimes of
Rydberg atoms~\cite{MeasurLifetime} and may cause undesirable
frequency shifts, for example in atomic clocks \cite{SuperOL}.

The source responsible for both virtual
and real photon exchange is the dipole-dipole interaction.
It governs the build-up as well as the decay of coherence
in a dense radiating sample.  On the one hand, the virtual
exchange of photons is responsible for the so-called exchange
interaction. Its strength is exemplified by the energy difference
$2\hbar \Omega=\wp^2/2\pi\epsilon_0r^3$ between the symmetric
and anti-symmetric single-excitation superposition
$\ket{\pm}=(\ket{eg}\pm\ket{ge})/\sqrt{2}$ of two atoms in
their ground $g$ or excited $e$ states separated by $r$.
On the other hand, the real photon exchange is responsible
for dephasing of a dense gas and has the same $r^{-3}$
dependence. The interplay of both determines whether
the decay speed-up in a dense inverted gas of two-level
atoms is mostly incoherent (intensity proportional to atom
number $N$, called ``amplified spontaneous emission'', ASE)
or coherent ($\propto N^2$, called ``superradiance'' or
``superfluorescence''). Experimentally, this difference can
be seen in whether there is an initial build-up in the decay
intensity, due to the $N^2$ dependence, or not.

The difficulty of calculating effects including atom-atom
cooperation relates to the intractably large number of
interconnected degrees of freedom, even if just a few particles
are involved. To explore these collective effects, many new ideas,
such as the quantum jump approach, were developed to treat
superradiance~\cite{QjumpSuper,QjumpSuper03,clemens}. Recently, we
successfully incorporated cooperative effects into a novel
formalism for optically dense media. The result is a two-atom
master equation for
superradiance~\cite{SusanneSuperForm,SusanneAtomAtom1,SusanneAtomAtom2}.
We apply our model of cooperative radiation build-up to explain
the results of an experiment measuring rapid decay of an ultracold
Rb Rydberg gas.

The model, as used in \cite{SusanneAtomAtom1,SusanneSuperForm}, is based on perturbation theory carried to second order in the strength of the exchange interaction. Thus, we can eliminate all field and most atomic degrees of freedom which results in an effective two-atom nonlinear equation of motion of the Linblad type,
\begin{eqnarray}
\label{eq:master}
    \dot{\rho} &=&
    - \frac{1}{2}\sum\limits_{i,j=1,2} \Gamma_{ij}
    \left( \comm{\rho\s_{i}}{\s_{j}^\dagger} +
      \comm{\s_{i}}{\s_{j}^\dagger\rho} \right)\nonumber\\
    &&
    -\frac{1}{2}\sum\limits_{i,j=1,2} (\Gamma_{ij}+\gamma\delta_{ij})
    \left( \comm{\rho\s_{j}^\dagger}{\s_{i}} +
      \comm{\s_{j}^\dagger}{\s_{i}\rho} \right),\nonumber
\end{eqnarray}
where $\rho$ is a two-atom density operator, $\sigma_i^{(\dagger)}$ is the lowering (raising) operator of the $i$th atom, $\gamma$ the spontaneous emission rate, and $\Gamma_{ij}$ contains the second order dipole-dipole interaction between atoms $i$ and $j$. (First order effects lead to local field effects which don't play a role here \cite{SusanneSuperForm}). In order to obtain this result, Gaussian (and therefore, classical) light field statistics are assumed, in line with the second order approximation. In addition, a Markov approximation is made which is justified if the coherence time of the light fields is shorter than the atomic evolution \footnote{This Markov approximation is justified self-consistently: the times for atomic and field evolution are compared in the result of the calculation. Although for the fastest evolution times it is not strictly correct to assume atomic evolution to be much slower than field evolution, the approximation is still expected to show good qualitative results. Quantitative estimates of the validity of this procedure will be presented in an upcoming publication.}. Atomic collisions and center-of-mass motion are neglected.

The $\Gamma_{ij}$ operators can be calculated from
$\Gamma_{ij}\delta(t-t')\propto\cum{E_i(t) E_j(t'}$,
where $E_i$ denotes the quantum field at the location of atom $i$,
and the cumulant $\cum{A B}\equiv\erw{A B}-\erw{A}\erw{B}$.
$\Gamma_{ij}$'s contain both the virtual and real photon exchange,
and can be calculated for different systems. They can be expressed
only as highly nonlinear and implicit functions of the atomic
variables $\rho$ (\eqs{eq:Gamma}).
For small enough probe diameters $d$ retardation effects can be neglected. This approximation is justified in our case because the time it takes for light to propagate through the sample ($\sim 10^{-10}$ s) is significantly shorter than any other time in the system, in particular, the atomic build-up time. Note that sample-sizes less than the cubic wavelength, as needed in the Dicke model \cite{Dicke} are not necessary. Thus we can set $\Gamma_{ii}\equiv\Gamma$ and $\Gamma_{ij\ne i}\equiv \bar\Gamma$ and simplify \eq{eq:master}:
\begin{subequations}
\label{eq:anxDyn}
\begin{eqnarray}
\dot{\rho}_{ee} &=& -(2\Gamma+\gamma) \rho_{ee} + \Gamma\;,\\
\dot{m} &=& -2(2\Gamma+\gamma) m - 2\gamma(2\rho_{ee}-1) + 8\bar{\Gamma} \rho_{egge} \;, \\
\dot{\rho}_{egge} &=& - (2\Gamma+\gamma) \rho_{egge} + \bar{\Gamma} m  \;.
\end{eqnarray}
\end{subequations}
The upper-level population is $\rho_{ee}$, the inversion product $m=(\rho_{ee}-\rho_{gg})^2$, and the two-atom non-diagonal coupling $\rho_{egge}={\rm Tr}\rho\ket{eg}\bra{ge}$. (Setting $\rho_{egge}=0$ would lead to the usual single-atom formalism.)
In addition, we use
\begin{eqnarray}
\Gm & = & \gm \, \frac{\rho_{ee}}{2\rho_{ee}-1} \, \left( \e^{2\zeta}-1
\right) + 2\gm\Cc^2\varrho^4 \, \frac{\gm}{\Gm+\gm/2} \, \rho_{egge} \, I\left(\zeta,\varrho\right) \nonumber\\
\label{eq:Gamma}
\bar{\Gm}  & = & 3 \gm\Cc\varrho \, \frac{\gm}{\Gm+\gm/2} \, \rho_{ee} \,
I\left(\zeta,\varrho\right) +\\
 && \qquad  2\gm\Cc^2\varrho^4 \,
\frac{\gm}{\Gm+\gm/2} \, \rho_{egge} \, I\left(\zeta,\varrho\right)\;\nonumber
\end{eqnarray}
where
\begin{eqnarray*}
\zeta = \frac{1}{2}
\Cc\varrho\frac{\gm}{\Gm+\gm/2} (2\rho_{ee}-1)&,&
I(\zeta,\varrho) = \left|
\frac{\e^\xi(1-\xi)+1}{\xi^2}\right|^2_{\xi=\zeta+i\varrho}.
\end{eqnarray*}
The sample size $\varrho=\pi d/\lambda$ is measured relative to the wavelength of the light.

In our initial experiment, we have studied the decay of high-$n$ states using a simple detection scheme with only limited state-specificity. First,
Rb atoms
were trapped and  cooled to 100 $\mu K$. Next, they were selectively
excited by a pulsed UV laser to the $40p$ state. After a delay time
$\tau$, all atoms in states with principal quantum numbers
$n\geq$ 27 were Stark ionized. The remaining experimental details are the same as in \cite{farooqi03}.
\begin{figure}[ht]
    \centerline{\includegraphics[clip,width=.95\linewidth]{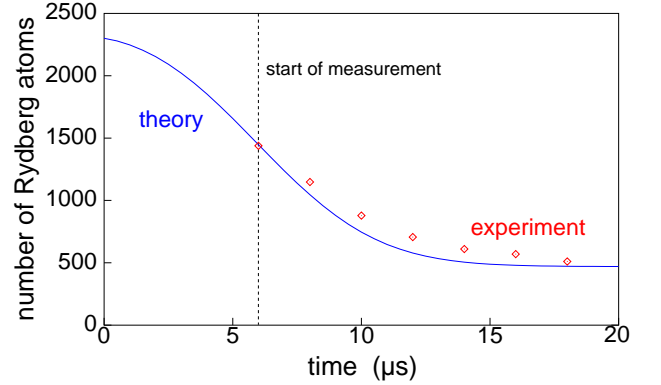}}
    \caption{Measured and calculated decay of the number of atoms in Rydberg states with $n\geq 27$ following excitation to $n=43p$. The initial density of Rydberg atoms in the experiment is $5\times 10^8$ cm$^{-3}$ The dots are experimental points, the solid line theoretical simulation. The fitting parameter in this calculation was the number of atoms present at the start of the measurement, i.e., 1400 Rydberg atoms at 6 $\mu$s.}
    \label{fig:SuperExp}
\end{figure}
As depicted in
Fig.~\ref{fig:SuperExp}, the number of ultracold atoms in Rydberg
states with $n\geq 27$ decays fast, an estimated 100 times faster than
expected in vacuum \footnote{This factor is a
conservative estimate at this time, and it will be quantified more
exactly in an improved experimental setting.}. We find that this speed-up can be explained by the presence of superradiance and, on some transitions, ASE
~\cite{RydSuper1,RydSuper2,RydSuper3}. (Alternative explanations for the strength of the speed-up would include so-called avalanche plasma formation \cite{robinson}, where a large fraction of the initial Rydberg atoms would be ionized. However, we rule this out because we measure only 190 free ions after a delay of 35 $\mu$s.)

In what follows, we will show that \eqs{eq:anxDyn} lead to excellent agreement with the experiment (see \fig{fig:SuperExp}). The density in the calculation is chosen to be the same as in the experiment, $5\times 10^8$ cm$^{-3}$. The sample in the experiment is cigar shaped, thus enabling good mode selection (as in all superradiance experiments to date). In the calculation we make the approximation of having, for each transition, only one mode, and then use, for calculational ease, a spherical geometry with the same sample volume as in the experiment.

\begin{figure}[ht]
    \centerline{\includegraphics[clip,width=.95\linewidth]{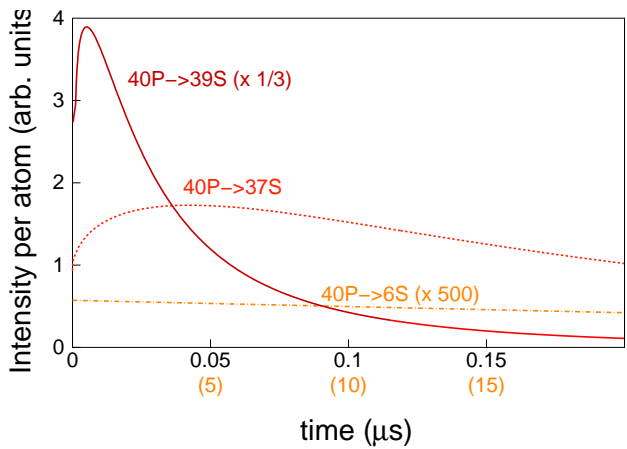}}
    \caption{Calculated
output intensity as function of time for a sample with density $5\times 10^8 cm^{-3}$ for the transition from state $40p$ to $39s$, $37s$, and $6s$, respectively.
The initial increase in intensity over time is the sign for superradiance, i.e., the decay into $39s$ and $37s$ qualifies as superradiant, whereas the decay into $6s$ does not. (The curve for $6s$ is shown on a 100 times faster timescale to show the decay.)}
    \label{fig:AandMaDot}
\end{figure}

\begin{figure}[ht]
    \centerline{\includegraphics[clip,width=.95\linewidth]{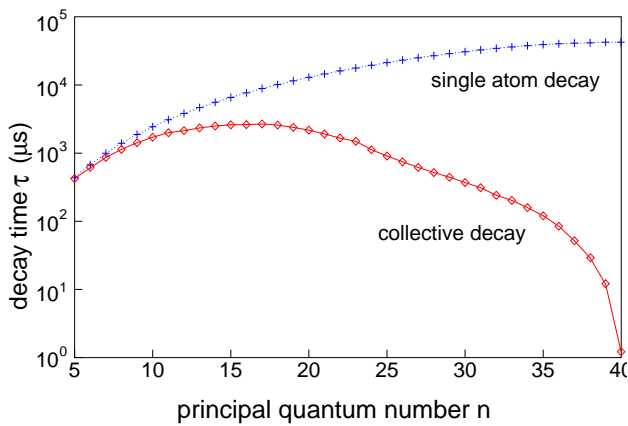}}
    \caption{Decay times from 40P to various $n$S states $(\diamond)$ in a dense gas ($\Nc =5*10^8\ cm^{-3}$) and in vacuum
    $(+)$. }
    \label{fig:SuperVsNon}
\end{figure}

The simulations presented here assume Rb atoms in the initial
state $40p$. In Fig.~\ref{fig:AandMaDot} we show the
decay from $40p$ into $ns$. In Fig.~\ref{fig:SuperVsNon}, the
effective decay times are compared for a dense gas and a
vacuum (cf.~\cite{vrinceanu}). In vacuum, the effective decay time $\tau_{\rm eff}$ is
the inverse of the Einstein A-coefficient. Clearly, in a vacuum the
transition into the states with lowest $n$ is fastest, and
therefore decay into these channels is by far the most likely. But this
tendency is reversed dramatically in dense gases: the effective decay time for
each transition is shorter by up to three orders of magnitude than that in a vacuum or in dilute gases.
Since the collective and cooperative effects responsible for this
speed-up depend only on the density relative to the wavelength cubed, the acceleration of the decay is obviously stronger
for longer wavelengths.
Figure~\ref{fig:SuperVsNon} and the quantitative form of the increase in decay for higher densities, particularly for low frequencies, are one of the main results presented in this letter.

In Fig.~\ref{fig:AandMaDot}, the intensity of some selective
decays is shown over
time. Because of energy conservation, the intensity must be
proportional to the negative time derivative of the upper state
population. (We neglect here all time-delay effects, resulting in
an instantaneous intensity at time $t=0$.)
In this graph, our (somewhat arbitrary) distinction between ASE and superradiance can be seen: An initially positive slope of intensity over time, as seen for $40p\To 39s$  is associated with superradiance, whereas a monotonically decreasing intensity, as seen for $40p\To 6s$ means ASE. It is important to emphasize here again that in reality there is no sharp boundary as there are coherent and incoherent elements mixed in all decays, thus making the transition between the two cases very smooth.

\begin{figure}[ht]
    \centerline{\includegraphics[clip,width=.95\linewidth]{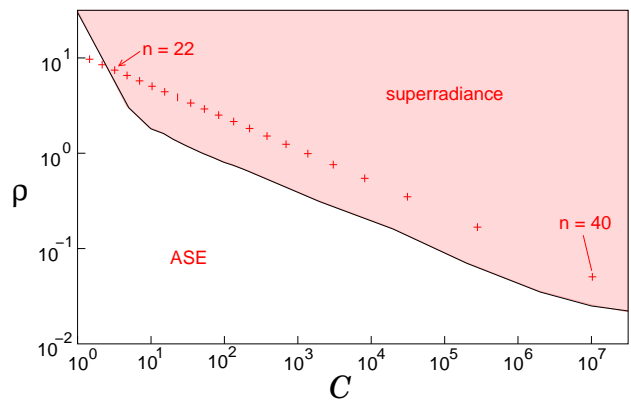}}
    \caption{Map of critical parameters of $\Cc$ and
    $\varrho$ $(\diamond)$. Above the critical curve (shaded area) are the parameters for which superradiance happens. Also shown are the $\Cc$ and
    $\varrho$ for the decay to $ns$ states from 40$p$ state $(+)$. Density of atoms is the same as above.}
    \label{fig:criticalCCrho}
\end{figure}

In order to get a general overview of which combination of
parameters leads to superradiance, we created a map in the $\Cc$--$\varrho$ parameter space with relative density or cooperative parameter $\Cc$ and relative size $\varrho$. Figure \nofig{fig:criticalCCrho} shows the numerically determined
 border, as defined above, between superradiant and ASE behavior.
The selective decay from the $40p$
Rydberg state of Rb into all possible lower $ns$ states is
added to the map. We see that superradiant behavior is expected
for decay into levels with $n\geq$ 22.

We discuss now the calculation shown in \fig{fig:SuperExp}. Theoretically, we can calculate, from the decay times as presented in \fig{fig:SuperVsNon}, the lifetime of $40p$ (and the lower states) directly,
\[
\frac{1}{\tau_{\rm total}}=\sum_{\parbox{2cm}{\rm\tiny all channels out of $40p$}}\frac{1}{\tau_{\rm eff}},
\]
and we find $\tau_{\rm total}\approx 5 \mu$s. This is to be compared to a $\tau_{\rm total}^{(0)}=210 \mu$s for dilute gas or vacuum. The experiment, however, cannot measure this time directly but only the total lifetime of all states with $n\ge 27$. In order to compare our theoretical method with the experiment we simulate a cascade from $40p$ via all intermediate states down to $n<27$, using the decay times in \fig{fig:SuperVsNon} and analogous times for the $p, d, f$, etc. states with $5\le n\le40$. This procedure is approximated by using only the two fastest channels out of each state. Numerically, we can compare this result with one that uses one channel more per state and find only  small changes of 1-10\%. The result is depicted in the strong black curve in \fig{fig:SuperExp}, which shows excellent agreement with the experiment.

Future experiments with improved state-selective detection will allow direct comparison to the single-lifetime calculations.

In this article, we have discussed the possibility
of superradiant decay in cold gases of Rydberg atoms at densities
of 10$^8$ -- 10$^9$ cm$^{-3}$. Superradiance occurs because
lower-frequency decays are increasingly more likely to happen in
denser gases, and they contribute most to cooperative behavior.
Level shifts due to atomic interactions may inhibit superradiance at higher densities and/or higher $n$. This could explain why superradiance is not routinely seen.

We have neglected black body radiation, since it is important for
superradiance only for the initiation of the radiation
process and only if  $N\gg n_B$, where $n_B$ is the average
number of black body photons per mode at the frequency of
transitions~\cite{RydSuperBlack}. In addition, the possibility of
mode competition and interference between different decay channels
is neglected for simplification. In future work, the effects of
geometry, in particular the aspect ratio of the sample, should be
taken into account. In practice, only elongated samples are used to show superradiance \cite{SuperRev}.

Using our calculation we were able to
obtain close agreement with observed signatures of superradiance
including the effects of dissipation and the unique temporal
build-up of a sharp flash of radiation. Moreover, our new
formalism allows for easy incorporation of more complicated level
structures, additional fields, and polarization effects.

In summary, recent experiments measuring the decay of ultracold Rb Rydberg atoms find rates much faster than that of atoms in dilute gases. These results are consistent with superradiant behavior in the framework of our
model.

The authors gratefully acknowledge support from the National
Science Foundation and the Research Corporation. DV wishes to thank DOE for support through the
Los Alamos National Laboratories. We want to thank J. Riccobono for discussions.

\bibliography{RydSuper}
\end{document}